\documentclass[pra,showkeys,showpacs,groupedaddress,twocolumn,longbibliography]{revtex4-1}
\usepackage{graphicx}
\usepackage{dcolumn}
\usepackage{bm}
\usepackage{color}
\usepackage[caption=false]{subfig}

\begin{document}
\title{An atomic receiver for AM and FM radio communication}

\author{D.~A.~Anderson$^{1}$}
\email{dave@rydbergtechnologies.com.}
\author{R.~E.~Sapiro$^{1}$}
\author{G.~Raithel$^{1,2}$}
\affiliation{1. Rydberg Technologies, Ann Arbor, MI 48104 USA}
\affiliation{2. Department of Physics, University of Michigan, Ann Arbor, MI 48109 USA}


\begin{abstract}
Radio reception relies on antennas for the collection of electromagnetic fields carrying information, and receiver elements for demodulation and retrieval of the transmitted information.  Here we demonstrate an atom-based receiver for AM and FM microwave communication with a 3-dB bandwidth in the baseband of $\sim$100~kHz that provides optical circuit-free field pickup, multi-band carrier capability, and inherently high field sensitivity. The quantum receiver exploits field-sensitive cesium Rydberg vapors in a centimeter-sized glass cell, and quantum-optical readout of baseband signals modulated onto carriers with frequencies ranging over four octaves, from C-band to Q-band.  Receiver bandwidth, dynamic range and sideband suppression are characterized, and acquisition of audio waveforms of human vocals demonstrated. The atomic radio receiver is a valuable receiver technology because it does not require antenna structures and is resilient against electromagnetic interference, while affording multi-band operation in a single compact receiving element.
\end{abstract}

\maketitle
\section{Introduction}
Since its advent in the late nineteenth century, radio communication in the audio band has been an integral component of society and proven essential to the advancement of science and technology~\cite{Edison.1878,Tesla2.1900,MarconiBraunnobel.1909}.
Ongoing challenges include information security, bandwidth increase by tapping the microwave- and mm-wave regimes~\cite{Rappaport.2013}, and resilience against electromagnetic interference (EMI). In recent years, quantum technologies have become a topic of interest in this field, as they exploit quantum-mechanical phenomena to build devices for secure data transfer~\cite{Hiskett.2006,Ursin.2007,Yin.2017}, realize small atom-based receivers that harness the high sensitivity of atoms~\cite{Barredo.2013,Sedlacek.2012,Holloway2.2014,Gordon.2014} and artificial quantum structures~\cite{Floess.2017, Zamani.2017, Peng.2014} to electromagnetic fields, and quantum receivers that employ the immunity of atoms to very intense fields~\cite{Anderson.2016,Anderson.2017} for wire-free EMI-tolerant detector technology. An emerging type of quantum sensor employs Rydberg atoms, which are atoms in highly excited atomic states~\cite{Gallagher,Frey.1993} that offer a giant electromagnetic response due to resonant frequency matching with radio-frequency and microwave radiation (RF), at carrier frequencies typically used in communications, ranging from tens of MHz into the THz range.  The atomic response arises from the strong electric-dipole coupling of the atoms to the electromagnetic field and can be retrieved quantum-optically using electromagnetically induced transparency (EIT)~\cite{Mohapatra.2007}, a spectroscopic method that enables optical measurement of RF fields, without circuitry required within the atom-based RF field sensing structure. The growing interest in Rydberg-atom-based field measurement includes the development of calibration-free, non-invasive sensors for static and time-varying electric~\cite{Barredo.2013,Sedlacek.2012,Holloway2.2014,Gordon.2014,Anderson.2016,Miller.2016,Anderson.2017,Jiao.2017} fields, as well as for magnetic fields up to 1~Tesla~\cite{Whiting.2016,Ma.2017,Zhang.2017}.

Pulse and analog modulation of carrier waves allows the adaptation of atomic sensors to applications in communications tasks. For instance, atom-based magnetic sensors have already been proposed for communication based on binary phase shift keying~\cite{Gerginov}, and digital communication based on Rydberg atoms has recently been realized~\cite{Meyer.2018}. In this work, we demonstrate a Rydberg quantum sensor to receive, play back and record baseband signals in the audio range that are amplitude-modulated (AM) or frequency-modulated (FM) onto microwave carriers~\cite{AndersonProvisional.2017}.  A single quantum detector is employed for multi-band AM and FM communication using carrier frequencies spanning more than four octaves, from the C-band to Q-band.  The atomic radio wave receiver (or ``atomic radio'') operates by direct real-time optical detection of the atomic response to AM and FM baseband signals, precluding the need for traditional de-modulation and signal-conditioning electronics. The small atomic vapor-cell sensor head replaces the function of traditional antennas, realizing a compact multi-band receiving element.  The absence of any circuitry within the receiver components exposed to the incident fields makes the atomic radio inherently EMI-tolerant. In the presented work we develop the principle of operation of the Rydberg atomic radio. We employ an experimental demonstration unit to characterize the baseband bandwidths and dynamic ranges for both AM and FM modulation, and we record audio samples of human vocals.  The demonstrated atomic radio exhibits good performance over the entire human audio band, with an upper baseband frequency limit exceeding 100~kHz.  We discuss the selection of laser operating points that maximize radio performance for given spectroscopic signatures, which depend on the atomic transitions and carrier waves used.  While the achieved dynamic range between 20~dB and 30~dB falls slightly short of radio standards, we see considerable room for improvement of our demonstration unit.

\section{Experimental setup}
\begin{figure}
\centering
\includegraphics[width=1\linewidth]{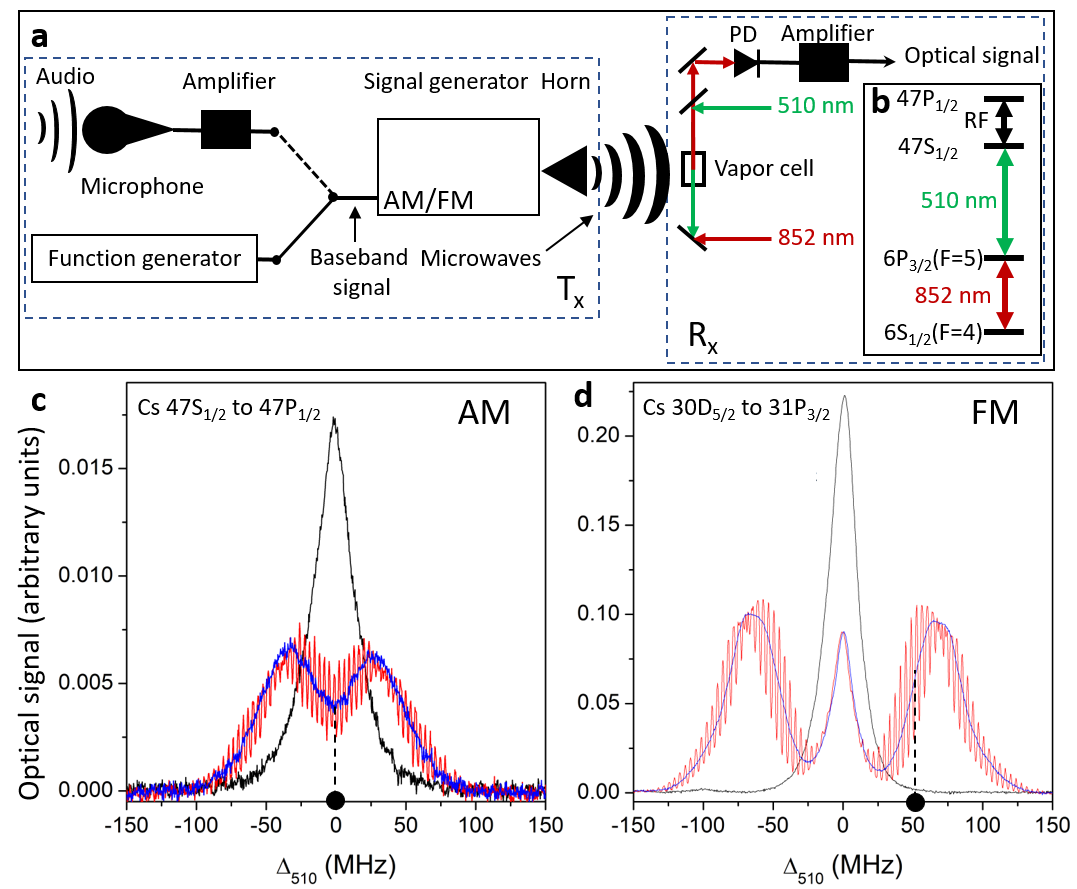}
\caption{\bf a \rm Experimental setup. \bf b \rm Energy-level diagram.  \bf c \rm Spectral readout of the Cs 47S$_{1/2}$ Rydberg line without microwave signal (black), with an un-modulated 37.4065~GHz carrier resonantly driving the Cs 47S$_{1/2}$ to 47P$_{1/2}$ Rydberg transition (blue), and with the carrier amplitude-modulated at baseband AM frequency 1~kHz and modulation depth $\pm 25\%$ (red). \bf d \rm Spectral readout of the Cs 30D$_{5/2}$ Rydberg line without microwave (black), with an un-modulated 29.458~GHz carrier resonantly driving the Cs 30D$_{5/2}$ to 31P$_{3/2}$ Rydberg transition (blue),
and with the carrier frequency-modulated at baseband FM frequency 1~kHz and modulation deviation $\pm 30$~MHz (red).}
\label{fig:figure1}
\end{figure}

The experimental setup comprising of the transmitter system and quantum RF receiver is illustrated in Fig.~\ref{fig:figure1}a.  The transmitter system consists of either an audio source collected by a microphone or a function generator that produces a baseband signal used to amplitude-modulate or frequency-modulate a microwave carrier.  The microwave signal, produced by a microwave signal generator, is fed into and emitted from a horn antenna, and is directed towards an atomic quantum RF receiver located several centimeters away.  The quantum receiver implements a 20~mm spectroscopic cell containing a cesium (Cs) vapor as the detector element.  The incident AM and FM transmission signals are detected by real-time optical readout of field-sensitive Rydberg states of the Cs vapor in the cell using electromagnetically induced transparency (EIT). An energy-level diagram for the Rydberg-EIT optical readout on the Cs 47S$_{1/2}$ Rydberg state is shown in Fig.~\ref{fig:figure1}b.  The EIT readout is performed using a probe beam with a wavelength of 852~nm focused to $330~\mu$m full-width-half-maximum (FWHM), that is counter-propagated and overlapped with a coupler beam with a wavelength of $\sim$510~nm focused to  $390~\mu$m FWHM.  The probe is frequency-stabilized to the Cs $6S_{1/2}~F=4\rightarrow6P_{3/2}~F'=5$ transition, and its absorption through the cell is monitored on a photodiode, while the coupling laser is set near the chosen $6P_{3/2}\rightarrow Rydberg$ transition.  The coupler-beam wavelength is tuned to reach the desired Rydberg state. Two experimental modes are then employed. To analyze the Rydberg EIT lines and to identify coupler-laser frequency operating points suitable for radio reception, the coupler frequency is scanned linearly at a rate of a few Hz over a few hundred MHz across the selected Rydberg transition (Figures~\ref{fig:figure1} and~\ref{fig:figure4}). In radio reception mode, the coupler frequency is held fixed at an operating point near the center of a Rydberg resonance; this mode is suitable for real-time baseband detection (Figures~\ref{fig:figure2},~\ref{fig:figure3}~and~\ref{fig:figure5}).  The probe photodiode signal is amplified by a preamplifier and then acquired by an oscilloscope, connected to an A-to-D converter for digital recording, or plugged directly into an audio speaker.

\section{Optical readout of AM or FM microwave carrier fields}

Atoms with a quasi-free electron in a high-lying Rydberg state exhibit large polarizabilities and electric dipole moments that scale with principal quantum number $n$ as $\sim n^7$ and $\sim n^2$, respectively, rendering them exquisitely sensitive to electric fields~\cite{Gallagher}. Figure~\ref{fig:figure1}c shows the spectroscopic EIT readout from the vapor-cell receiver of the Cs 47S$_{1/2}$ Rydberg line without RF applied (black curve) and with application of a 37.4065~GHz K$_{\rm a}$-band un-modulated carrier wave resonant with the Cs 47S$_{1/2}$ to 47P$_{1/2}$ Rydberg transition at a power of -5~dBm injected into the transmission line and horn (blue curve).  Application of the RF field causes the line to split symmetrically into a pair of Autler-Townes (AT) lines whose separation corresponds to the Rabi frequency $\Omega$ of the Rydberg/RF-field interaction.  The measured splitting is directly proportional to the microwave carrier electric field strength and is given by the relation $E=\hbar\Omega/d$, where $d$ is the transition dipole moment of the Rydberg transition.  From the observed splitting in Fig.~\ref{fig:figure1}c, we measure a carrier field amplitude at the location of the receiver atoms of $E$=5.9~V/m.

A Rydberg-EIT-AT spectrum taken with a microwave carrier that is AM-modulated by a sinusoidal 1~kHz baseband signal with $\pm 25\%$ modulation depth is overlaid in Fig.~\ref{fig:figure1}c (red curve).  The laser detuning $\Delta_{510}$ is scanned linearly at a rate of $\sim$10~Hz across the atomic resonance.  The 1~kHz-AM of the carrier manifests in the optical sampling/readout in the form of periodic oscillations that occur while $\Delta_{510}$ is slowly scanned across the EIT-AT resonance. The oscillations result from the dependence of the Rydberg-EIT-AT spectrum on the RF electric field amplitude. The AM causes a modulation of the AT-splitting of the spectrum and of the corresponding optical signal, which occurs rapidly and concurrently with the relatively slow scan of $\Delta_{510}$ across the AT-split 47S$_{1/2}$ to 47P$_{1/2}$ resonance. The vertical range of the AM-induced excursions of the optical signal provides a measure for the AM-sensitivity of the Rydberg-EIT-AT signal as a function of $\Delta_{510}$. In the case displayed in Fig.~\ref{fig:figure1}c (red curve), the detector exhibits a maximal AM response at $\Delta_{510}=0$, the center of the 47S$_{1/2}$ to 47P$_{1/2}$ Rydberg-EIT-AT resonance. Hence, for the case in Fig.~\ref{fig:figure1}c  $\Delta_{510}=0$ is chosen as the coupler-laser frequency operating point for real-time detection of AM baseband signals (black dot and dashed line in  Fig.~\ref{fig:figure1}c).

Rydberg states of atoms and their resonant AT responses to RF fields are, in addition to being dependent on RF amplitude, also susceptible to changes in RF field frequency~\cite{Berman}.  In the weak-field limit of resonant AT, a change in the RF field frequency manifests in an asymmetry of the AT-line-pair signal strength as well as in a variation of the AT-splitting and the AT line positions. The combination of these effects affords baseband signal detection in FM-modulated carriers via a similar spectroscopic readout as with AM. Figure~\ref{fig:figure1}d shows the optical readout from the receiver of the 30D$_{5/2}$ atomic Rydberg line without RF applied (black curve), with an un-modulated 29.458~GHz K$_{\rm a}$-band 6.3-V/m carrier wave that is resonant with the Cs 30D$_{5/2}$ to 31P$_{3/2}$ Rydberg transition (blue curve), and with a 1~kHz sinusoidal baseband signal FM-modulated onto the carrier at a $\pm$30~MHz maximal deviation (red curve).  For D to P transitions, dipole selection rules dictate that linearly-polarized microwave fields may only couple $\Delta m_j =0$ Rydberg transitions.  As a result, the 30D$_{5/2}$ to 31P$_{3/2}$ $m_j$=1/2 and 3/2 components are respectively coupled by the microwave field and exhibit similar AT splittings, whereas the 30D$_{5/2}$ $m_j=5/2$ level that is not coupled to another state remains unaffected by the RF and produces a line centered at $\Delta_{\rm 510}=0$.  It follows that the desired coupler-laser operating frequency for FM baseband detection using resonant D to P Rydberg transitions is different from that when using S to P transitions.  This is apparent in Fig.~\ref{fig:figure1}d,  where we see that for D to P transitions the atomic response to the FM signal is maximal at the inner inflection points of the AT-split pair (black dot and dashed line in  Fig.~\ref{fig:figure1}d).

\section{Dynamic range and harmonic response to AM and FM signals.}
\begin{figure}
\centering
\includegraphics[width=1\linewidth]{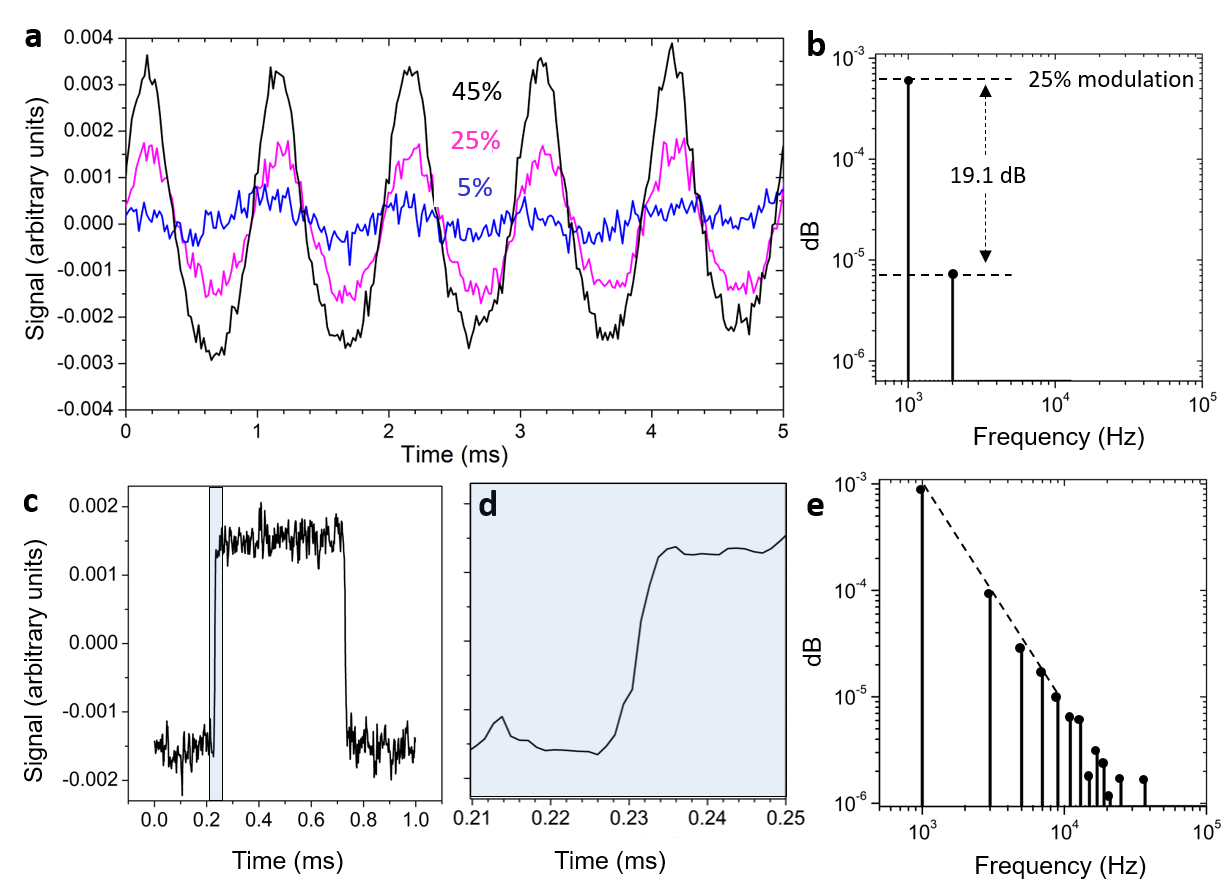}
\caption{Optical response to AM with harmonic and square-wave baseband signals on the 47S$_{1/2}$ to 47P$_{1/2}$ Rydberg transition. \bf a \rm Real-time optical readout for a 1~kHz sinusoidal baseband signal with AM modulation depths of $\pm$5\% (blue), $\pm$25\% (purple), and $\pm$45\% (black) at laser operating point $\Delta_{510}$=0~MHz (see Fig.~\ref{fig:figure1}c). \bf b \rm Power spectrum of the received AM 1~kHz signal at a modulation depth of $\pm$25\%. \bf c \rm  Optical readout of a square pulse, \bf d \rm a zoom-in on its rising edge, and \bf e \rm its power spectrum.}
\label{fig:figure2}
\end{figure}

For application in communication systems, the baseband bandwidth and transient response behavior of the quantum receiver are useful performance metrics and benchmarks for comparison to traditional antenna-receivers.  In Fig.~\ref{fig:figure2} we characterize the response and detection speed of the quantum receiver for an AM-modulated 37.4~GHz carrier resonantly driving the 47S$_{1/2}$ to 47P$_{1/2}$ transition (Fig.~\ref{fig:figure1}b).  Here, we set $\Delta_{510}$=0~MHz at the field-free Rydberg resonance for maximum sensitivity to the AM field (Fig.~\ref{fig:figure1}b) and monitor the optical 852~nm transmission through the receiver cell in real time.  Figure~\ref{fig:figure2}a shows the optical readout of a harmonic 1~kHz baseband signal for a series of AM modulation depths over a 5~ms interval.  The measured optical readout maintains a high baseband signal visibility and integrity over a range of modulation depths from 5\% to 45\%, indicating a well-behaved, linear atomic response over a wide dynamic range.  To quantify the noise characteristics of the receiver we compute the power spectrum for the received AM signal at $\pm 25\%$ modulation depth (Fig.~\ref{fig:figure2}b).  The analysis reveals a single overtone at 2~kHz (second harmonic) that is 19.1~dB weaker than the fundamental. The overtone is attributed to the mildly nonlinear dependence of the Rydberg-EIT-AT signal at $\Delta_{510}$=0~MHz on the carrier amplitude.

To characterize the receiver's transient response, we transmit and receive a signal that is AM-modulated with a 1~kHz square-wave pulse in the baseband, which contains strong higher-harmonic content at odd overtone frequencies (Figs.~\ref{fig:figure2}c, d, e). The rise time of the received pulse is measured to be a few microseconds, indicating an upper limit to the receiver AM bandwidth of a few hundred kHz. The response time is limited by the bandwidth of the photodiode-amplifier used in our demonstration setup; the fundamental response time of the Rydberg-EIT-AT signal itself is shorter.  The power spectrum of the received square-wave pulse (Fig.~\ref{fig:figure2}c) is observed to consist of odd higher harmonics at $(2k+1)f$, with integer $k$ and fundamental AM frequency $f$, with spectral power decreasing as $dB\sim -2\log (k)$, as one would expect for an ideal square wave pulse (dashed line in Fig.~\ref{fig:figure2}e).

\begin{figure}
\centering
\includegraphics[width=1\linewidth]{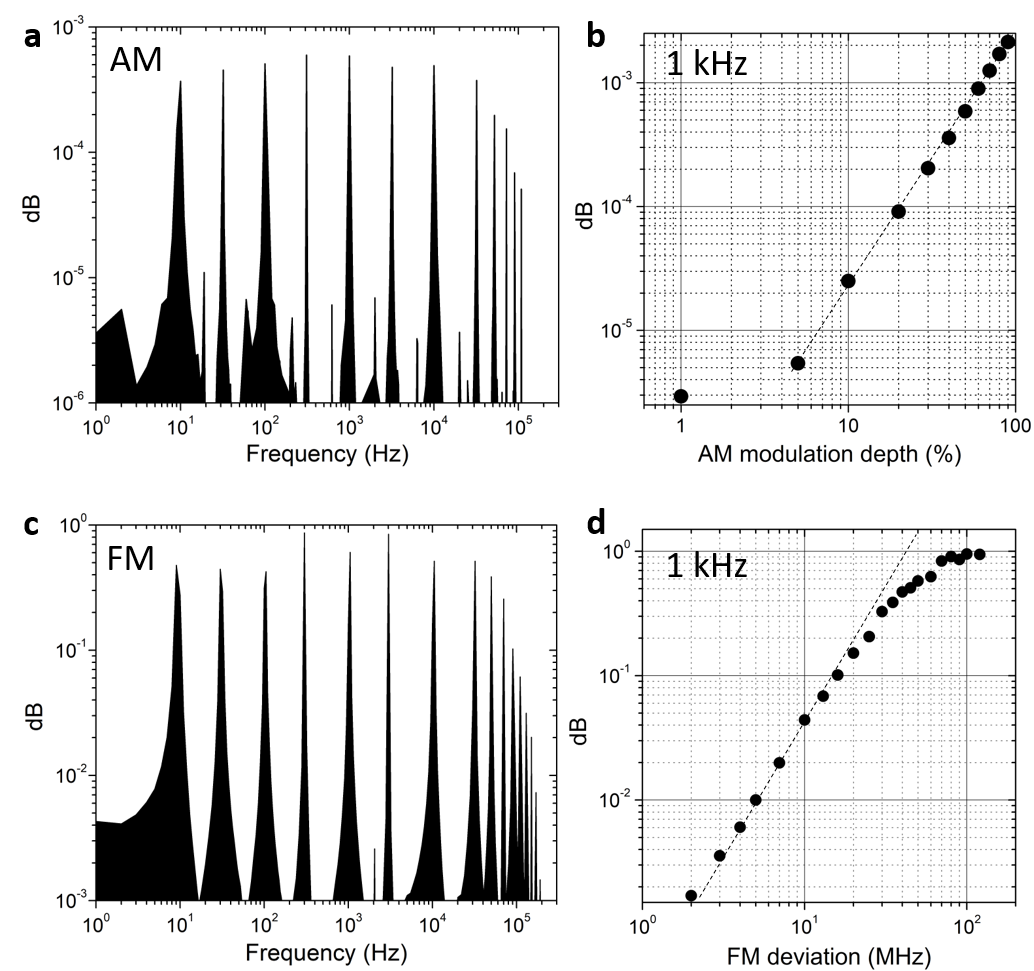}
\caption{Power spectral density and dynamic range of the quantum receiver for \bf a, b \rm  AM transmission on a 37.4~GHz carrier tuned to the 47S$_{1/2}$ to 47P$_{1/2}$ Rydberg transition with $\Delta_{510}=0$~MHz and \bf c, d \rm FM transmission on a 29.548~GHz carrier tuned to the 30D$_{5/2}$ to 31P$_{3/2}$ Rydberg transition with $\Delta_{510}\approx 50$~MHz.}
\label{fig:figure3}
\end{figure}

The harmonic response and dynamic range of the quantum receiver for both AM and FM transmission with $K_a$-band microwave carriers using the resonant Rydberg transitions and coupler-frequency operating points shown in Figs.~\ref{fig:figure1}c and d are presented in Figure~\ref{fig:figure3}.  In Figs.~\ref{fig:figure3}a and~c  we plot the measured power spectral density at a selection of baseband modulation frequencies for AM and FM modulation, respectively; these figures reveal the baseband bandwidth of the Rydberg radio. For both AM and FM, the measured power spectral densities indicate a 3-dB baseband bandwidth $\gtrsim 100$~kHz. In Figs.~\ref{fig:figure3}b and~d we plot the power of the signal received at 1~kHz modulation frequency as a function of AM modulation depth and FM deviation, respectively; these figures show the dynamic range (at 1~kHz).  For the experimental conditions used in the present work, including selected Rydberg transitions, carrier strengths, and coupler-laser-frequency operating points, the receiver exhibits a dynamic range between 20 and 30~dB, with sensitivities reaching AM modulation depths of a few percent and FM deviations of a few hundred kHz.

\section{Multi-band operation and audio recording}
\begin{figure}
\centering
\includegraphics[width=1\linewidth]{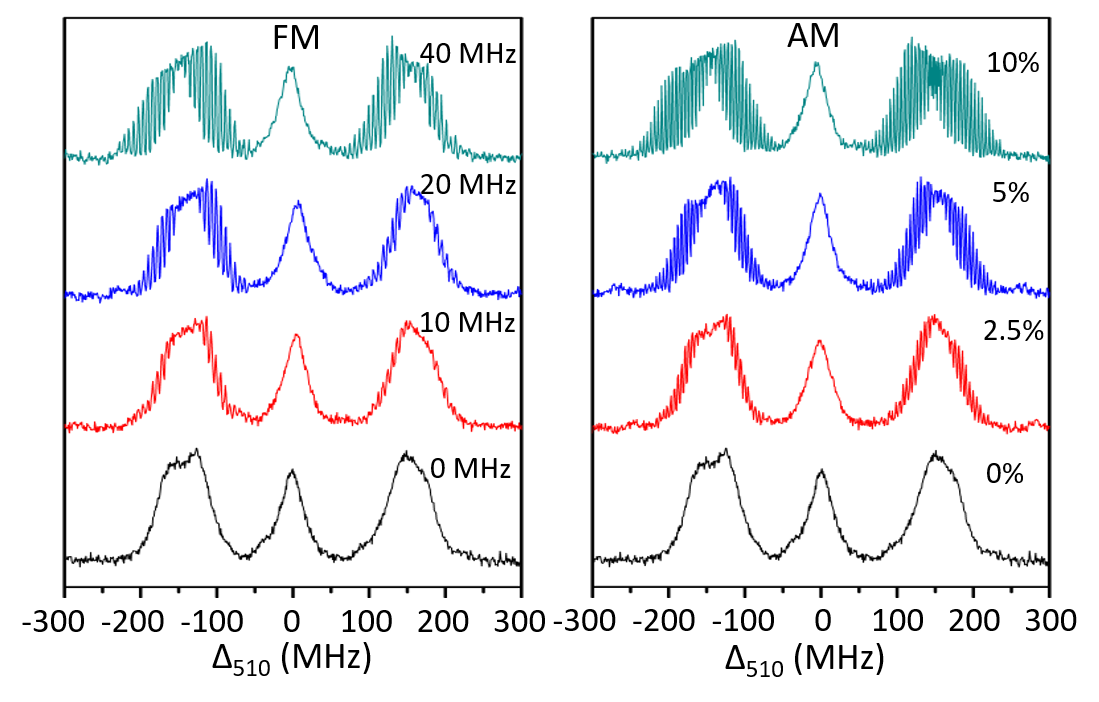}
\caption{Optical spectral readout of a 1~kHz baseband signal at a 4.5~GHz C-band RF carrier frequency for FM (left) and AM (right) transmission using the resonant Cs 54D$_{5/2}$ to 55P$_{3/2}$ Rydberg transition.  Spectra corresponding to a series of FM deviations and AM modulation depths are shown, as indicated.  The spectra are centered to the RF-field-free Cs 54D$_{5/2}$ Rydberg line.}
\label{fig:figure4}
\end{figure}

The wide selection of field-sensitive atomic Rydberg transitions that can be optically accessed using EIT allows a single vapor-cell quantum RF receiver to operate over an unprecedented range of carrier frequencies, from MHz to THz~\cite{Holloway2.2014}.  To demonstrate the multi-band capability, we employ the quantum receiver for detection of AM and FM audio on a C-band carrier by tuning the frequency of the coupling laser to target a suitable Rydberg state.  In Fig.~\ref{fig:figure4} we show AT-EIT spectra centered on the 54D$_{5/2}$ Rydberg line with 1~kHz AM and FM baseband signals transmitted on a 4.5~GHz C-band carrier field resonant with the Cs 54D$_{5/2}$ to 55p$_{3/2}$ Rydberg transition for comparable series of AM modulation depths and FM deviations.  The dipole moment of the C-band transition used here is larger than those of the $K_a$-band transitions, leading to a higher field sensitivity and a slightly wider baseband dynamic range in the C-band case. The demonstrated robust performance in the C-band shows that the receiver covers multiple carrier bands, with the carrier frequency range spanning more than four octaves.

\begin{figure}
\centering
\includegraphics[width=1\linewidth]{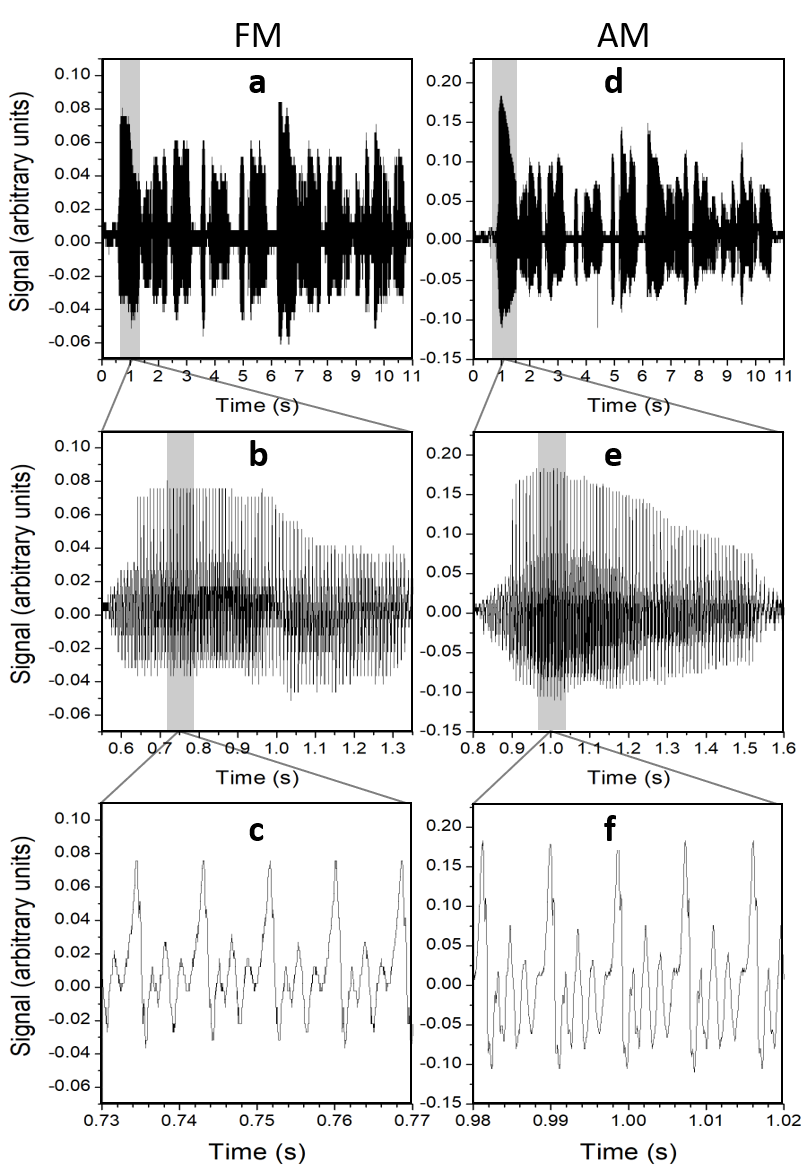}
\caption{Audio waveforms of human vocals recorded from received FM (\bf a, b, c\rm) and AM (\bf d, e, f\rm) audio transmission at a 4.5~GHz (C-band) carrier.}
\label{fig:figure5}
\end{figure}

Inspection of the strongly modulated spectra in Fig.~\ref{fig:figure4} also reveals a difference in EIT-AT response for AM and FM detection that may arise near the peaks of the AT-line pairs.  For FM detection, the readout near the center of the AT peaks generally exhibits a minimum in the observable modulation.  This is due to a cancellation of the simultaneous change in relative AT-peak signal strength and the AT line shift.  For AM detection, on the other hand, the modulated AT-splitting at the peak leads to a doubling of the optically-detected modulation signal-frequency.  The identification of these features in the present demonstration supports the choice of coupler-frequency operating points away from the AT-split peak centers.  Nevertheless, these features may prove advantageous to realizing advanced operation modes of the quantum receiver.

For a demonstration of the full functionality of the quantum receiver in radio communications in the audio band, we implement our unit for AM and FM C-band transmission recording and playback of human vocals.  Recorded waveforms of a human voice singing the first verse of 'Mary had a little Lamb' transmitted via AM and FM on a 4.5~GHz carrier and received by the quantum sensor are shown in Fig.~\ref{fig:figure5}.  Higher-resolution plots of equivalent sub-sections of the full original waveforms reveal details of the complex audio waveforms captured by the receiver, and variations in harmonic content and volume between the two separate AM and FM recordings are reproduced.  For audio files of recordings using the atomic receiver please contact the authors.

\section{Conclusion}
An atom-based quantum AM and FM receiver with a 3-dB bandwidth in the baseband of $\sim$100~kHz has been demonstrated.  The receiver is based on direct atom-optical detection and demodulation of AM and FM baseband signals on carriers ranging from C-band to Q-band, with a single quantum sensing element.  We have investigated the selection of laser operating points that maximize radio performance for the given spectroscopic signatures, which depend on the atomic transitions and carrier-wave fields used. We have characterized the harmonic response and measured a dynamic range between 20~dB and 30~dB for our demonstration unit.  The presented quantum receiver is based on direct quantum-optical collection and demodulation of AM and FM radio waves and paves the way for radio communication applications inaccessible to existing antenna-receiver technology, such as multi-carrier-frequency operation combined with high field-sensitivity in a single, compact detector unit.  Further development of Rydberg-based quantum receivers towards higher-bandwidth data transfer over a wider range of carrier frequencies appears readily accessible.  Upon completion of this manuscript we became aware of related work on optical detection of modulated carrier signals using Rydberg atoms~\cite{Deb.2018}.

\section{Acknowledgement}
This work was supported by Rydberg Technologies.


%

\end{document}